\documentclass[aps, pra, reprint, showpacs, amsmath, amssymb]{revtex4-1}

\usepackage{amsmath,amscd}
\usepackage{bm}% bold math
\usepackage[english]{babel}
\usepackage{graphicx}% Include figure files
\usepackage{soul}
\usepackage[usenames]{color}
\DeclareMathOperator{\sgn}{\textrm{sgn}}

\begin{document}
\title{ Cherenkov Radiation of a Charge Flying through the ``Inverted'' Conical Target }

\author{ Andrey V. Tyukhtin }
\email{a.tyuhtin@spbu.ru}
\author{ Sergey N. Galyamin }
\author{ Viktor V. Vorobev }
\author{ Aleksandra A. Grigoreva }
\affiliation{Saint Petersburg State University, 7/9 Universitetskaya nab., St. Petersburg, 199034 Russia}

\date{\today}

\begin{abstract}
Radiation generated by a charge moving through a vacuum channel in a dielectric cone is analyzed. 
It is assumed that the charge moves through the cone from the apex side to the base side (the case of ``inverted'' cone). The cone size is supposed to be much larger than the wavelengths under consideration. We calculate the wave field outside the target using the ``aperture method'' developed in our previous papers. Contrary to the problems considered earlier, here the wave which incidences directly on the aperture is not the main wave, while the wave once reflected from the lateral surface is much more important. The general formulas for the radiation field are obtained, and the particular cases of the ray optics area and the Fraunhofer area are analyzed. Significant physical effects including the phenomenon  of ``Cherenkov spotlight'' are discussed. 
In particular it is shown that this phenomenon allows reaching essential enhancement of the radiation intensity in the far-field region at certain selection of the problem parameters. Owing to the ``inverted'' cone geometry, this effect can be realized for arbitrary charge velocity, including the ultra relativistic case, by proper selection of the cone material and the apex angle. 
Typical radiation patterns in the far-field area are demonstrated. 
\end{abstract}

\maketitle

\section{Introduction\label{sec:intro}}

Cherenkov radiation (CR) produced by a moving charged particle in various complicated targets was extensively studied several decades ago in the context of development of Cherenkov detectors and counters~%
\cite{Jb, Zrb}%
.
Mentioned targets (or, more specifically, radiators) were typically dielectric (solid or liquid) objects like rods, cones, prisms, spheres or their combinations.
Proper manipulation with the emitted radiation (mainly for focusing purposes) was typically performed by external mirrors and lenses or, less frequently, by specific form and coating of the radiator surfaces.
For example, a cylindrical radiator with the external conical mirror was utilized in the first experiments by P.~Cherenkov~%
\cite{Ch37}%
.
Later on, conical radiators with flat or spherical end surfaces (or rods with the conical or spherical end) were considered~%
\cite{Getting47, Dicke47, Marshall51, Marshall52}%
.
The idea to form the optical surface so that the CR may be focused at a single stage of reflexion or refraction has been also discussed~%
\cite{Jb}%
.
Moreover, similar conical and prismatic targets were investigated in the context of development of radiation sources in microwave region based on CR effect~%
\cite{Danos53, Coleman60}%
.

In recent years, the renewed interest to the aforementioned objects has emerged.
The main applications of interest are development of beam-driven radiation sources (based on high-quality beams produced by modern accelerators) and non-invasive systems for bunch diagnostics.
For example, both prismatic target and hollow conical target with the flat out surface (accompanied by the set of external mirrors) have been used in a series of experiments on microwave and Terahertz CR~%
\cite{Takahashi00, Sei15, Sei17}%
.
The papers~%
\cite{Sei15, Sei17}
should be especially noted in the context of the present paper since they used similar radiator.
A high-power Terahertz source based on dielectric cone having its apex facing the incident electron beam has been proposed in~%
\cite{Sei15}
while the paper~%
\cite{Sei17}
has demonstrated the first experimental results on generation of coherent CR by such a target at Kyoto University linac.
A prismatic radiator (similar to that used in~%
\cite{Danos53}%
) was proposed for CR-based bunch diagnostics in~%
\cite{Pot10}%
.
Later on, a similar prismatic target with one reflecting flat surface was discussed as a prospective candidate for simultaneous monitoring of electron and positron beams at CESR storage ring~%
\cite{Bergamproc17}%
.
Corresponding experimental results showing the prominent possibilities of this scheme have been reported in recent paper~%
\cite{Kieffer18}%
.

For further development of the discussed topics, an efficient and reliable approach is needed for analytical investigation of CR field generated by charged particle bunches in various dielectric radiators of complex shape.
Historically, various approaches, different from paper to paper, were utilized for this purpose.
For example, analytical description of CR from complicated radiators of Cherenkov detectors was typically performed using the CR theory in infinite medium (Tamm-Frank theory~%
\cite{TF37, T39}%
) and simple ray optics laws~%
\cite{Jb,Zrb}%
.
In the papers~%
\cite{Takahashi00, Sei15, Sei17}
the interaction between the charge and the boundary of the target closest to the charge trajectory was taken into account semi-analytically.
Similar approach (taking into account only the internal target's boundary) was used in~%
\cite{Coleman60}
for calculation of total radiated CR energy from hollow conical radiator.
In the paper~%
\cite{Kieffer18}%
, an exponential decay in CR intensity with an increase in the impact parameter was calculated using the so-called polarization current approximation~%
\cite{Pot10}%
.
However, all the mentioned analytical approaches do not take into account all the essential properties of radiators and the produced CR.

Starting from the paper~%
\cite{BTG13}%
, we are developing two combined approaches which take into account both the internal radiator's surface (which is mainly interacting with the charged particle bunch) and the out radiator's surface (through which the CR goes into free space)~%
\cite{BTG13, BGT15, GT14, GTAB14, GTV18, TVBG18, TVGB19, TGV19}%
.
Moreover, one of these approaches (the ``aperture approach'') allows correct calculation of the CR field in the far-field (Fraunhofer) zone and near caustics formed by convergent rays where ray optics fails~%
\cite{GT14, GTAB14, GTV18, TVBG18, TVGB19, TGV19}%
.
It should be underlined that though some distinct parts of these approaches were discussed and utilized earlier, their proper combinations were not collected into convenient analytical methods.
It is also equally important that our approaches were successfully verified via wave simulations in COMSOL~%
\cite{GTV18, TVBG18, TGV19}%
.
Below we briefly explain the main steps of our methods.

Two first steps of these methods are the same.
At the first step, CR field in the bulk of the target is calculated.
We suppose that this field is the same as in the corresponding ``etalon'' problem, while the latter is the problem with the medium having only the inner boundary, i.e. the boundary closest to the charged particle trajectory.
It is also imposed that the ``etalon'' problem has an analytical solution.
For example, for radiators with the flat surface, this is the problem with a charge moving along the plane interface between two media.
Known solution of this ``etalon'' problem~%
\cite{TdF60, Ulrich66, B62}
was utilized in~%
\cite{Takahashi00}
and our papers~%
\cite{BGT15, TVGB19}%
.
For radiators having a cylindrical channel, this is a problem of a charge passing through a hole in an infinite dielectric medium with the solution given in~%
\cite{BBol57, B62, GVT19, GVT19E}%
.
This solution was utilized in~%
\cite{Coleman60}
and our papers~%
\cite{BTG13, GT14, GTV18, TVBG18, TGV19}%
.
It is worth noting that since the ``etalon'' problem is solved rigorously, arbitrary impact parameters or channel radii (including those of order of wavelength
$ \lambda $%
) can be considered.

At the second step, we return the out boundary and select the part of it illuminated by CR (we call this part an ``aperture'' and sign it as
$ \Sigma $%
).

We assume that the radiator is large, i.e. (i)
$ \sqrt{ \Sigma } \gg \lambda $
and (ii) the distance from the charge trajectory to
$ \Sigma $
is large compared to
$ \lambda $%
.
These assumptions allows considering CR at
$ \Sigma $
in the form of asymptotic being the quasi-plane wave (with small cylindrical wave front curvature).
This wave can be decomposed into two orthogonal polarizations.
Further the Snell and Fresnel laws can be used
for calculation of the field at the outer side of the aperture.

The third step is different for two methods being developed.
The ray-optics method uses the ray-optics laws (including those accounting for ray tube transformation~%
\cite{Fockb, KravOrb}%
) for calculation of the wave field outside the object~%
\cite{BTG13, BGT15}%
.
However, this technique has essential limitations.
First, the so-called ``wave parameter'' should be small,
$ D \sim \lambda L / \Sigma \ll 1 $
(%
$ L $
is the distance from
$ \Sigma $
to the observer), this means that we cannot consider the important Fraunhofer area where
$ D \gg 1 $%
.
Moreover, the observation point cannot be in the neighborhood of focuses and caustics, where ray optics is not applicable.

The aperture method utilizes Stratton-Chu formulas (also frequently called the ``aperture integrals'') to calculate the field outside the target~%
\cite{GT14, GTAB14, GTV18, TVBG18, TVGB19, TGV19}%
.
This approach is lacking additional limitations of the ray-optics method and allows calculating the CR field both at caustics and focuses and at arbitrary distance
$ L $
corresponding to
$ D \sim 1 $
or
$ D \gg 1 $
(Fraunhofer or far-field area).

This paper is devoted to the study of CR produced by single charged particle (or a charged particle bunch) moving along the axis of the dielectric cone with the vacuum channel in the configuration similar to that in~%
\cite{Getting47, Sei15, Sei17}
(i.e. with the cone apex facing the incident charge).
Throughout this paper, we will refer to this geometry as the ``inverted'' cone to clearly distinguish between this case and analogues ``ordinary'' conical target with its base facing the incident charge which was analyzed in our previous papers~%
\cite{BTG13, TGV19}%
.
In particular, we have studied the phenomena of ``Cherenkov spotlight'' resulting in significant enhancement of CR intensity in the far-field zone~%
\cite{TGV19}%
.
However, this valuable effect takes place for certain strict limitations for the charge velocity and the cone angle only.
As we will show below, in the ``inverted'' configuration considered here, corresponding conditions are much simpler to fulfill, and therefore this effect is more attractive for practical realization. 
It should be noted that in this paper we mainly use the aperture method (since it is more general), however ray-optics solution is also derived as a specific case using the saddle point approach.

The paper is organized as follows.
Section~%
\ref{sec:apintgen}
briefly recalls the basic Stratton-Chu formulas and its simplified form for the far-field area.
Section~%
\ref{sec:etalon}
contains solution of the ``etalon'' problem and calculation of the CR field on the aperture.
The form of the aperture integrals for the problem under consideration is given in Sec.~%
\ref{sec:apint}%
, the particular case of ray-optics area is considered in Sec.~%
\ref{sec:rayoptics}
while the Fraunhofer area is considered in Sec.~%
\ref{sec:fraunhofer}%
.
Section~%
\ref{sec:spotlight}
is devoted to the detailed analysis of the ``Cherenkov spotlight'' regime.
Section~%
\ref{sec:numeric}
present the typical graphical results, while Sec.~%
\ref{sec:concl}
finishes the paper.

%%%%%%%%%%%%%%%%%%%%%%%%%%%%%%%%%%%%%%%%%%%%%%%%%%%%%%%%%%%%%%%%%
%%%%%%%%%%%%%%%%%%%%%%%%%%%%%%%%%%%%%%%%%%%%%%%%%%%%%%%%%%%%%%%%%
\section{\label{sec:apintgen} Aperture integrals: general form and approximation for Fraunhofer zone }
%%%%%%%%%%%%%%%%%%%%%%%%%%%%%%%%%%%%%%%%%%%%%%%%%%%%%%%%%%%%%%%%%
%%%%%%%%%%%%%%%%%%%%%%%%%%%%%%%%%%%%%%%%%%%%%%%%%%%%%%%%%%%%%%%%%

Aperture integrals (or Stratton-Chu formulae~%
\cite{Fradb}%
) for Fourier transform of electric field can be written in the following general form (we use Gaussian system of units)~%
\cite{TVBG18,TVGB19,TGV19}%
:
\begin{equation}
\begin{aligned}
& \vec{ E } \left( \vec{ R } \right)
=
\vec{ E }^{ ( h ) }
\left( \vec{ R } \right)
+
\vec{ E }^{ ( e ) }
\left( \vec{ R } \right), \\
& \vec{ E }^{ ( h ) }
\left( \vec{ R } \right)
=
\frac{ i k }
{ 4 \pi }
\int\limits_{ \Sigma }
\left\{
\left[
\vec{ n^{ \prime } } \times \vec{ H } \left( \vec{ R^{ \prime } } \right)
\right]
G \left(
\left|
\vec{ R } - \vec{ R^{ \prime } }
\right|
\right)
+
\right. \\
& \left. +
\frac{ 1 }{ k^2 }
\left(
\left[ \vec{ n^{ \prime } } \times \vec{ H }
\left( \vec{ R^{ \prime } } \right)
\right] \cdot \nabla^{ \prime }
\right)
\nabla^{ \prime }
G \left( \left|
\vec{ R } - \vec{ R^{ \prime } }
\right|
\right)
\right\}
d\Sigma^{ \prime }, \\
& \vec{E}^{ ( e ) }
\left( \vec{ R } \right)
= \\
& \frac{ 1 }{ 4 \pi }
\int\limits_{ \Sigma }
\left[
\left[
\vec{ n^{ \prime } }\times \vec{ E }
\left( \vec{ R^{ \prime } } \right)
\right] \times
\nabla^{ \prime }
G \left(
\left| \vec{ R } - \vec{ R^{ \prime } } \right|
\right)
\right] d\Sigma^{ \prime },
\end{aligned}
\label{eq:(1)}
\end{equation}
where
$ \Sigma $
is an aperture area,
$ \vec{ E }( \vec{ R^{ \prime } } ) $,
$ \vec{ H }( \vec{ R^{ \prime } } ) $
is the field on the aperture,
$ k = \omega / c = 2 \pi / \lambda $
is a wave number of the outer space (vacuum),
$ \lambda $
is a wavelength under consideration,
$ \vec{ n^{ \prime } } $
is a unit external normal to the aperture at the point
$\vec{ R^{ \prime } } $,
$ G( R ) = \exp ( i k R ) / R $
is a Green function of the Helmholtz equation, and
$ \nabla^{ \prime } $
is a gradient:
$ \nabla^{ \prime } = \vec{e}_x \partial / \partial x^{ \prime } + \vec{e}_y \partial  / \partial y^{ \prime } + \vec{e}_z \partial / \partial z^{ \prime } $.
Analogous formulas are known for the magnetic field as well, however we do not write them here because we are mainly interested in the ``wave'' zone
$ k L \gg 1 $
(%
$ L $
is a distance from the aperture to the observation point) where
$\left| \vec{E} \right| \approx \left| \vec{H} \right| $.

The observation point is often located in the region called the Fraunhofer area (far-field area) where so-called ``wave parameter''
$ D $
is large:
\begin{equation}
D
\sim
\lambda R / \Sigma
\sim
\lambda R / d^2 \gg 1,
\label{eq:(2)}
\end{equation}
where
$ \lambda $
is a wavelength under consideration,
$ R $
is a distance from the target to the observation point, and
$ \Sigma \sim d^2 $
is a square of an aperture (we assume that the origin of the coordinate frame is located in the vicinity of the target; in this case
$ R \sim L$%
).
It is of interest to simplify the general formulae~%
\eqref{eq:(1)}
in this area.

Note that the condition~%
\eqref{eq:(2)}
automatically results in the inequality
\begin{equation}
R \gg d \cdot d / \lambda \gg d,
\label{eq:(3)}
\end{equation}
because
$ d \gg \lambda $
in the problem under consideration.
Using the inequalities~%
\eqref{eq:(2)},
\eqref{eq:(3)}
and taking into account that
$ \left| \vec{R}^{ \prime } \right| \sim d $
one can apply the following approximation in the formulae~%
\eqref{eq:(1)}:
\begin{equation}
G\left( \left| \vec{R}-\vec{R}' \right| \right)
\approx
\frac{
\exp
\left(
i k R - i k \vec{R} \vec{R}^{ \prime } / R
\right)}
{R}.
\label{eq:(4)}
\end{equation}
As a result, we obtain the following formulae for the Fraunhofer area:
\begin{widetext}
\begin{equation}
\begin{aligned}
& \vec{ E }^{ ( h ) }
\left( \vec{ R } \right)
\approx
\frac{ i k \exp ( i k R ) }{ 4 \pi R }
\int\limits_{ \Sigma }
\left\{
\left[
\vec{ n^{ \prime } } \times \vec{ H } \left( \vec{ R^{ \prime } } \right)
\right]
-
\vec{ e }_R \left( \vec{ e }_R \cdot
\left[ \vec{ n^{ \prime } } \times \vec{ H } \left( \vec{ R^{ \prime } } \right)
\right]
\right)
\right\}
\exp \left( -i k \vec{ e }_R \vec{ R^{ \prime } } \right)
d\Sigma^{ \prime } , \\
& \vec{ E }^{ ( e ) }
\left( \vec{ R } \right)
\approx
\frac{ i k \exp \left( i k R \right) }{ 4 \pi R }
\int\limits_{ \Sigma }
\left[
\vec{ e }_R \times
\left[ \vec{ n^{ \prime } } \times \vec{ E }
\left( \vec{ R^{ \prime } } \right)
\right]
\right]
\exp \left( -i k \vec{ e }_R \vec{ R^{ \prime } } \right)
d \Sigma^{ \prime },
\end{aligned}
\label{eq:(5)}
\end{equation}
\end{widetext}
where
$ \vec{e}_R = \vec{ R } / R $.
The formulae~%
\eqref{eq:(5)}
can have essential advantages in comparison with~%
\eqref{eq:(1)}
for specific objects because we can hope to evaluate these integrals analytically.

%%%%%%%%%%%%%%%%%%%%%%%%%%%%%%%%%%%%%%%%%%%%%%%%%%%%%%%%%%%%%%%%%
%%%%%%%%%%%%%%%%%%%%%%%%%%%%%%%%%%%%%%%%%%%%%%%%%%%%%%%%%%%%%%%%%
\section{\label{sec:etalon} The field on the aperture }
%%%%%%%%%%%%%%%%%%%%%%%%%%%%%%%%%%%%%%%%%%%%%%%%%%%%%%%%%%%%%%%%%
%%%%%%%%%%%%%%%%%%%%%%%%%%%%%%%%%%%%%%%%%%%%%%%%%%%%%%%%%%%%%%%%%

\begin{figure}
\centering
\includegraphics[width=0.85\linewidth]{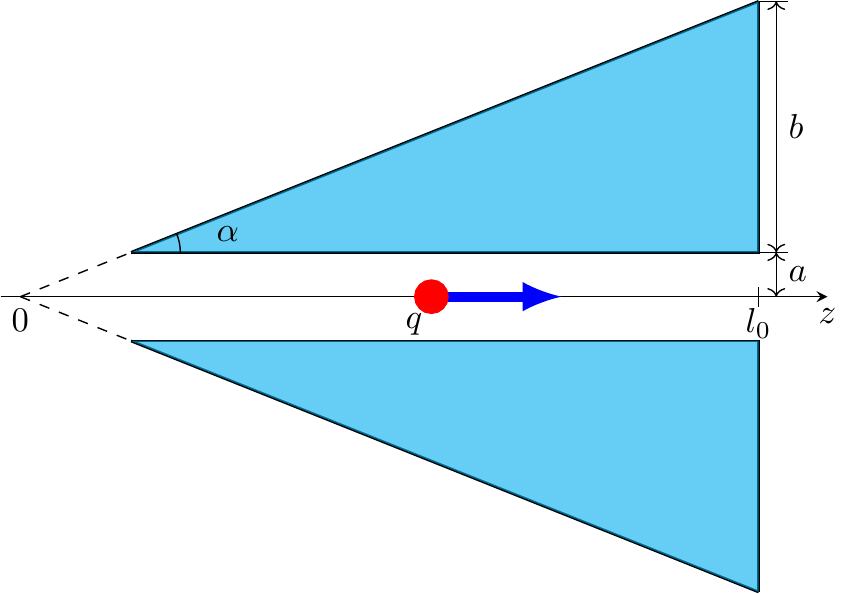}
\caption{\label{fig:1}%
The cone cross-section.
}
\end{figure}

We analyze radiation of a charge moving along the axis of the cylindrical channel with radius
$ a $
in a conical object (Fig.~%
\ref{fig:1}%
).
The target is made of a material with permittivity
$ \varepsilon $
and permeability
$ \mu $
(the conductivity is assumed to be negligible).
The width of the ring at the cone base is
$ b $
(the radius of the cone base is
$ b + a $%
), and the cone angle is
$ \alpha $.
Accordingly, the length of the target along its axis is
$ l = b \cot \alpha $,
and the distance from the top of the cone to its base is
$ l_0 = ( a + b ) \cot \alpha $.
The target sizes are much larger than the wavelength under consideration:
$b \gg \lambda $
and
$ l \gg \lambda $.
The coordinate system origin is at the cone apex, and the
$ z $%
-axis is the symmetry axis of the target.

The charge
$ q $
moves with constant velocity
$ \vec{ V } = c \beta \vec{ e }_z $
along the
$ z $%
-axis into the cone from the apex side.
For definiteness, we will deal with a point charge having the charge density
$ \rho = q \delta ( x ) \delta ( y ) \delta ( z - V t ) $,
where
$ \delta $
is a Dirac delta-function.
However, the results obtained below can be easily generalized for the thin bunch with finite length because we consider Fourier transforms of the field components.

It is assumed that the charge velocity exceeds the ``Cherenkov threshold'', i.e.
$ \beta > 1 / n $,
where
$ n = \sqrt{ \varepsilon \mu }$
is a refractive index of the target material.
Thus, CR is generated in the cone material.
We suppose that the generatrix cone surface reflects CR with the the coefficient
$ R_{ v0 } $%
.
Similarly, the end surface of the cone transmits radiation with the transmission coefficient
$ T_v $
which can be calculated by the Fresnel formulas
(the index
$ v $
indicates that the field has ``vertical'' polarization).
Note that we are interested in the case when the radiation does not experience total internal reflection at the cone end surface.

We write first the ``initial'' incident field that is the field in the infinite medium with the channel~%
\cite{TVBG18, TGV19, B62}.
This field has the vertical polarization with non-zero components
$ H_{ \varphi }^{ i 0 } $,
$ E_r^{ i 0 } $,
$ E_z^{ i 0 } $
(cylindrical coordinate system
$ r $,
$ \varphi $,
$ z $
is used).
The Fourier-transform of the magnetic component at the distance
$ r \gg \lambda $
is
\begin{equation}
H_{ \varphi }^{ i 0 }
\approx
\frac{ q }{ c }
\eta
\sqrt{ \frac{ s }{ 2 \pi r} }
\exp
\left[ i \left( s r + \frac{ \omega }{ V } z - \frac{ \pi }{ 4 } \right) \right],
\label{eq:(6)}
\end{equation}
where
\begin{equation}
\begin{aligned}
&\eta
=
-\frac{ 2 i }{ \pi a } \times \\
& \times
\left[
\kappa \frac{ 1 {-} n^2 \beta^2 }{ \varepsilon \left( 1 {-} \beta^2 \right)}
I_1
\left( \kappa a \right)H_{0}^{ ( 1) }( s a ) {+} s I_0 ( \kappa a ) H_{1}^{ ( 1 ) }( s a )
\right]^{-1},
\end{aligned}
\label{eq:(7)}
\end{equation}
$ s ( \omega ) = \frac{ \omega }{ V } \sqrt{ n^2 \beta^2 - 1 } $,
$ \kappa ( \omega ) = \frac{ \left| \omega  \right| }{ V } \sqrt{ 1 - \beta^2 } $,
$ I_{ 0, 1 } $
are modified Bessel functions,
$ H_{ 0, 1 }^{ ( 1 ) } $
are Hankel functions.
Note that
$\mathrm{ Im } s ( \omega ) \ge 0 $
if we take into account a small dissipation.
If dissipation tends to zero then this condition results in the rule
$ \sgn \left[ s ( \omega ) \right] = \sgn ( \omega ) $.
%(we exclude the exotic case of so-called ``left-handed'' medium where
%$ \sgn \left[ s ( \omega ) \right] = - \sgn ( \omega ) $%
%).
The result~%
\eqref{eq:(6)}
is valid for
$ \left| s \right| r \gg 1 $.
The electric field
$ \vec{ E }^{ i 0 } $
can be easily found because the vectors
$ \vec{ E }^{ i 0 } $,
$ \vec{ H }^{ i 0 } $
and the wave vector of CR
$ \vec{ k }_{ i 0 } = s \vec{ e }_r + \vec{ e }_z \omega / V $
form the right-hand orthogonal triad in this area:
$ \vec{ E }^{ i 0 } = - \sqrt{ \mu / \varepsilon } \left[ \vec{ k }_{ i 0 } / k_{ i 0 }, \vec{ H }^{ i 0 } \right] $.
The angle between the wave vector
$ \vec{ k }_{ i 0 } $
and the charge velocity
$ \vec{ V } $
is
$ \theta_p = \arccos \left[ 1 / \left( n \beta \right) \right] $.

In accordance with the aperture method, we need to know the field which falls at the target's boundary being the ``aperture'' for the outer vacuum region.
In the case under consideration, the aperture is the part of the cone end surface which is illuminated by CR.

First, we need to take into account the Cherenkov wave, which directly falls on the base of the cone (it can be called the ``first'' wave).
The aperture for this wave is the entire base area.
This wave falls on the base at the Cherenkov angle
$ \theta_p $
and is refracted at the angle
$ \theta_{ t 1 } $
with the refraction coefficient
$ T_{ v 1 } $:
\begin{equation}
\begin{aligned}
\theta_{ i 1 }
&=
\theta_p, \\
\theta_{ t 1 }
&= \arcsin \left( n\sin \theta_p \right)
= \arcsin
\left(
\frac{ \sqrt{ n^2 \beta^2 - 1 } }{ \beta }
\right),
\end{aligned}
\label{eg:(8)}
\end{equation}
\begin{equation}
\begin{aligned}
T_{ v 1 }
&=
2 \sqrt{ \frac{ \mu }{ \varepsilon } }
\frac{ \cos \theta_{ i 1 } }
{ \sqrt{ \mu / \varepsilon } \cos \theta_{ i 1 }
+
\cos \theta_{ t 1 }
}
= \\
&=
2 \sqrt{ \frac{ \mu }{ \varepsilon } }
\frac{ 1 }
{ \sqrt{ \mu / \varepsilon } + n \sqrt{ 1 - \beta^2 ( n^2 - 1 ) } }.
\end{aligned}
\label{eq:(9)}
\end{equation}
Note that the effect of total internal reflection for this wave takes place under the condition
$  \beta^2 ( n^2 - 1 ) > 1 $.

Using Eq.~%
\eqref{eq:(6)}
it is easily to obtain the following expressions for the field of the first wave on the external surface of the cone base (at the point
$ r = r^{ \prime } $,
$ \varphi = \varphi^{ \prime } $,
$ z = l_0 + 0 $%
):
\begin{equation}
\begin{aligned}
&\left.
H_{ \varphi^{ \prime } }^{ ( 1 ) } \right|_{ z^{ \prime } = l_0 + 0 }
\approx
{ Q_1 } \frac {\exp ( i s r^{ \prime } )} { \sqrt{ k r^{ \prime } } }
={ Q_1 } \frac {\exp ( i k r^{ \prime } \sin \theta_{t1} ) } { \sqrt{ k r^{ \prime } } }   , \\
&E_{ r^{ \prime } }^{ ( 1 ) }
\approx
H_{ \varphi^{ \prime } }^{ ( 1 ) }
\cos \theta_{ t 1 },
\quad
E_{ z^{ \prime } }^{ ( 1 ) }
\approx
- H_{ \varphi^{ \prime } }^{ ( 1 ) }
\sin \theta_{ t 1 },
\end{aligned}
\label{eq:(10)}
\end{equation}
where
\begin{equation}
Q_1
=
\frac{ q k \eta \sqrt[4]{ n^2 \beta^2 - 1 } }{ c \sqrt{ 2 \pi \beta } }
T_{ v 1 }
\exp \left( \frac{ i k l_0 }{ \beta } - \frac{ i \pi }{ 4 } \right).
\label{eq:(11)}
\end{equation}
Note that we took into account in~%
\eqref{eq:(10)}
that the field under consideration is a quasi-plane transverse wave on the almost whole aperture.

However the first wave~%
\eqref{eq:(10)}%
, probably, is not a main wave in the important area where the angle
$ \theta $
is not large.
Indeed, for this wave, the angle of incidence
$\theta_{ i 1 } = \theta_p $,
as a rule, is not small, and
$ \theta_{ t 1 } > \theta_{ i 1 } $.
Therefore, we can expect that this wave will make a significant contribution only for not small angles
$\theta $.
To describe the radiation close to the
$ z $%
-axis, it is necessary to take into account the wave reflected from the lateral face of the cone (for brevity, we will call it the ``second wave''; it is shown in Fig.~%
\ref{fig:2}%
).
This wave can have small and even zero angles of incidence
$ \theta_{ i 2 } $
and refraction
$ \theta_{ t 2 } $.
Therefore this wave can be the main one in the region of relatively small angles
$ \theta $.

The initial wave~%
\eqref{eq:(6)}
falls on the lateral cone boundary at the angle
$
\theta_{ i 0 } = \pi / 2 + \alpha - \theta_p
$
(Fig.~%
\ref{fig:2}%
).
It is reflected at the same angle
$
\theta_{ r 0 } = \theta_{ i 0 }
$
and refracted at the angle
$ \theta_{ t 0 } = \arcsin ( n \sin \theta_{ i 0 } )$
with respect to the boundary normal (Fig.~%
\ref{fig:2}%
).
The wave reflected from the lateral surface is the wave which incidents on the cone base (%
$ \vec{ H }^{ i 2 } $%
).
It is a cylindrical wave, as the wave~%
\eqref{eq:(6)}.
We can write it at the point with cylindrical coordinates
$ r^{ \prime } $,
$ z^{\prime } $
in the following form:
\begin{equation}
H_{ \varphi }^{ i 2 }
\approx
\frac{ q }{ c }
R_{ v 0 } \eta \sqrt{ \frac{ s }{ 2 \pi r^{ \prime } } }
\exp \left[ i \Phi_{ i 2 } ( r^{ \prime }, z^{ \prime } ) \right],
\label{eq:(12)}
\end{equation}
where
\begin{equation}
R_{ v 0 }
=
\frac{ \sqrt{ \mu / \varepsilon } \cos \theta_{ i 0 } - \cos \theta_{ t 0 } }
{ \sqrt{ \mu / \varepsilon } \cos \theta_{ i 0 } + \cos \theta_{ t 0 } }
\label{eq:(13)}
\end{equation}
is the reflection coefficient from the lateral cone surface, and
$ \Phi_{ i 2 } ( r^{ \prime }, z^{ \prime } )$
is the phase which consists of two summands:
\begin{equation}
\Phi_{ i 2 } ( r^{ \prime }, z^{ \prime } )
=
\Phi_{ i 0 } ( r^{ \prime }, z^{ \prime } ) + \Delta \Phi_i ( r^{ \prime }, z^{ \prime } ).
\label{eq:(14)}
\end{equation}
Here
$ \Phi_{i0} ( r^{ \prime }, z^{ \prime } ) $
is the phase of the initial incident wave~%
\eqref{eq:(6)}
on the lateral surface, and
$ \Delta \Phi_i ( r^{ \prime }, z^{ \prime } ) $
is the additional phase acquired after reflection.

For the further calculation, we need to find the point of reflection from the lateral surface.
It is the solution of system of equation for the cone generatrix and the reflected ray equation:
\begin{equation}
r = z \tan \alpha,
\quad
r = r^{ \prime } + ( z - z^{ \prime } ) \tan \theta_{ i 2 },
\label{eq:(15)}
\end{equation}
where
\begin{equation}
\theta_{ i 2 }
=
\theta_{ r 0 } - \left( \pi /  2 - \alpha \right)
= 2 \alpha - \theta_p
\label{eq:(16)}
\end{equation}
is the angle of incidence at the cone base (it can be easily found from Fig.~%
\ref{fig:2}%
).
The solution of the system~%
\eqref{eq:(15)}
is
\begin{equation}
r_{ * } = z_{ * } \tan \alpha,
\quad
z_{ * } = \frac{ r^{ \prime } - z^{ \prime } \tan \theta_{ i 2 } }{ \tan \alpha - \tan \theta_{ i 2 } }.
\label{eq:(17)}
\end{equation}
Therefore, in accordance with~%
\eqref{eq:(6)}%
, the phase of the initial incident wave is
\begin{equation}
\begin{aligned}
&\Phi_{i 0 } ( r^{ \prime }, z^{ \prime } ) =
s r_{ * } + \frac{ \omega }{ V } z_{ * } - \frac{ \pi }{ 4 } = \\
&=
k n \cot( \theta_p - \alpha )
\left[ r^{ \prime } \cos \theta_{ i 2 } - z^{ \prime } \sin \theta_{ i 2 } \right] - \pi / 4.
\end{aligned}
\label{eq:(18)}
\end{equation}

Additional phase
$ \Delta \Phi_i ( r^{ \prime }, z^{ \prime } ) $
is equal to the product of the wave number in the medium (i.e.
$ k n $%
) by the length of the ray:
\begin{equation}
\Delta \Phi_i ( r^{ \prime }, z^{ \prime } )
=
k n
\frac{ z^{ \prime } - z_{ * } }{ \cos \theta_{ i 2 } }
= k n
\frac{ z^{ \prime } \sin \alpha - r^{ \prime } \cos \alpha }{ \sin ( \theta_p - \alpha ) }.
\label{eq:(19)}
\end{equation}
Summing up~%
\eqref{eq:(18)}
and
\eqref{eq:(19)}%
, after simple transformation we obtain
\begin{equation}
\begin{aligned}
& \Phi_{ i 2 } ( r^{ \prime }, z^{ \prime } )
=
k n \left( r^{ \prime } \sin \theta_{ i 2 } + z^{ \prime } \cos \theta_{ i 2 } \right)- \pi / 4
= \\
&=
k r^{ \prime } \sin \theta_{ t 2 } + k n z^{ \prime } \cos \theta_{ i 2 } - \pi / 4,
\end{aligned}
\label{eq:(20)}
\end{equation}
where
$ \theta_{ t 2 } $
is the angle of refraction of the 2-nd wave on the end surface of the cone (Fig.~%
\ref{fig:2}%
):
\begin{equation}
\theta_{ t 2 }
=\mathrm{ arcsin }( n \sin \theta_{ i 2 } ).
\label{eq:(21)}
\end{equation}

Using~%
\eqref{eq:(12)}
and
\eqref{eq:(20)}
one can obtain the Fourier-transform of the field on the external surface of the aperture (in the point with cylindrical coordinates
$ r^{ \prime } $,
$ \varphi^{ \prime } $,
$ z^{ \prime } = l_0 + 0 $%
) in the following form:
\begin{equation}
\begin{aligned}
\left. H_{ \varphi^{ \prime } }^{ ( 2 ) } \right|_{ z^{ \prime } = l_0 + 0 }
&\approx
Q_2
\frac{ \exp \left( i k r^{ \prime } \sin \theta_{ t 2 } \right) }{ \sqrt{ k r^{ \prime } } }, \\
E_{ r^{ \prime } }^{ ( 2 ) }
&\approx
H_{ \varphi^{ \prime } }^{ ( 2 ) }
\cos \theta_{ t 2 }, \\
E_{ z^{ \prime } }^{ ( 2 ) }
&\approx
- H_{ \varphi^{ \prime } }^{ ( 2 ) }
\sin \theta_{ t 2 },
\end{aligned}
\label{eq:(22)}
\end{equation}
where
\begin{equation}
\begin{aligned}
Q_2
&=
\frac{ q }{ c } \sqrt{ \frac{ k s }{ 2 \pi } }
R_{ v 0 }
T_{ v 2 }
\eta
e^{ i k n l_0 \cos \theta_{ i 2 } - i \pi / 4 }
= \\
&=
\frac{ q k \sqrt[4]{ n^2 \beta^2 - 1 } }
{ c \sqrt{ 2 \pi \beta } }
R_{ v 0 } T_{ v 2 }
\eta
e^{ i k n l_0 \cos \theta_{ i 2 } - i \pi / 4 },
\end{aligned}
\label{eq:(23)}
\end{equation}
and
$ T_{v2} $
is coefficient of refraction:
\begin{equation}
T_{ v 2 }
=
\frac{ 2 \sqrt{ \mu / \varepsilon } \cos \theta_{ i 2 } }
{ \sqrt{ \mu / \varepsilon } \cos \theta_{ i 2 } + \cos \theta_{ t 2 } }.
\label{eq:(24)}
\end{equation}

%%%%%%%%%%%%%%%%%%%%%%%%%%%%%%%%%%%%%%%%%%%%%%%%%%%%%%%%%%%%%%%%%%%%%%%%%%%%%%%%%%%%
\section{\label{sec:apint} Aperture integrals for the ``inverted'' cone }
%%%%%%%%%%%%%%%%%%%%%%%%%%%%%%%%%%%%%%%%%%%%%%%%%%%%%%%%%%%%%%%%%%%%%%%%%%%%%%%%%%%%

%
%%%%%%%%%%%%%%%%%%%%%%%%%%%%%%%%%%%%%%%%%%%%%%%%%%%%%%%%%%%%%%%%%%%%%%%%%%%%%%%%%%%%%%%%%
%%%%%%%%%%%%%%%%%%%%%%%%%%%%%%%%%%%%%%%%%%%%%%%%%%%%%%%%%%%%%%%%%%%%%%%%%%%%%%%%%%%%%%%%%%%%

Now we should write the general Stratton-Chu formulas~%
\eqref{eq:(1)}
in the form which is convenient for further calculation in the case of considered target.
Because of axial symmetry of the problem we can place the observation point in the plane
$ y = 0 $,
then
$ \vec{ e }_r = \vec{ e }_x $,
$ \vec{ e }_{ \varphi } = \vec{ e }_y $.
As well, we take into account that the normal to the aperture coincides with
$ z $%
-axis:
$ \vec{ n }^{ \prime } = \vec{ e }_z $.
We will use further the following formulas:
\begin{equation}
\tilde{ R } { = } \left| \vec{  R} {-} \vec{ R }^{ \prime } \right|
{=}
\sqrt{ r^2 {+} { r^{ \prime } }^2 {-} 2 r r^{ \prime } \cos \varphi^{ \prime } {+} ( z - l_0 )^2 },
\label{eq:(25)}
\end{equation}
\begin{equation}
\left[ \vec{ n }^{ \prime } \times \vec{ H }( \vec{ R }^{ \prime } ) \right]
=
- H_{ \varphi^{ \prime } } ( \vec{ R }^{ \prime } )
\left( \vec{ e }_r \cos \varphi^{ \prime } + \vec{ e }_{ \varphi } \sin \varphi^{ \prime } \right),
\label{eq:(26)}
\end{equation}
\begin{equation}
\nabla^{ \prime }
G( \tilde{R} ) =
\left(
\vec{ e }_{ r^{ \prime } } \partial_{ r^{ \prime } } +\frac{ \vec{ e }_{ \varphi^{ \prime } } }{ r^{ \prime } } \partial_{ \varphi^{ \prime } } + \vec{ e }_z \partial_{ z^{ \prime } }
\right)
G( \tilde{ R } ),
\label{eq:(27)}
\end{equation}
\begin{equation}
\begin{aligned}
\left[ \vec{ n }^{ \prime } \times \vec{ E }( \vec{ R }^{ \prime } ) \right]
&=
\left[ \vec{ e }_z \times \vec{ e }_{ r^{ \prime } } \right]\vec{ H }_{ \varphi^{ \prime } } ( \vec{ R }^{ \prime } )
\cos \theta_{tm}
= \\
&=
\vec{ e }_{ \varphi^{ \prime } } \vec{ H }_{ \varphi^{ \prime } }( \vec{ R }^{ \prime } ) \cos \theta_{tm},
\end{aligned}
\label{eq:(28)}
\end{equation}
\begin{equation}
\begin{aligned}
\left[
\left[
\vec{ n }^{ \prime } \times \vec{ E } ( \vec{ R }^{ \prime } ) \right]
\times \nabla^{ \prime }
\right]
&G( \tilde{ R } )
=
\vec{ H }_{ \varphi^{ \prime } } ( \vec{ R }^{ \prime } )
\cos \theta_{tm} \times \\
&\times
\left[
\vec{ e }_{ r^{ \prime } } \partial_{ z^{ \prime } } -\vec{ e }_z \partial_{ r^{ \prime } }
\right]
G( \tilde{ R } ),
\end{aligned}
\label{eq:(29)}
\end{equation}
where 
$ m $ 
is the number of the wave exiting the target ($ m = 1, 2 $).
Here, for brevity, we introduce the notation for the partial derivative:
$ \partial_x \equiv \partial / \partial x $
(further, analogously, the second derivative is written in the form
$ \partial_{ x y } \equiv \frac{ \partial^2 }{ \partial x \partial y }$%
).
	
Using~%
\eqref{eq:(25)} 
and 
\eqref{eq:(29)}, 
after a series of cumbersome transformations, one can obtain from~%
\eqref{eq:(1)}
the following result for the
$ m $%
-th part of the field generated by the wave with number
$ m $%
:

\ 

\begin{widetext}
\begin{equation}
\begin{aligned}
& E_{ \varphi }^{ ( m ) } = 0, \\
& \left\{
\begin{aligned}
& E_r^{ ( m ) }  \\
& E_z^{ ( m ) } \\
\end{aligned}
\right\}
=
-
\frac{ i }{ 2 \pi k }
\int\limits_{ r_{ m l } }^{ r_{ m h } }
dr^{ \prime }
\int\limits_{ 0 }^{ \pi }
d \varphi^{ \prime }
r^{ \prime }
 H_{ \varphi^{ \prime } }^{ ( m ) }( \vec{ R }^{ \prime } ) \times  \\
& \times
\left\{
\begin{aligned}
& k^2 \cos \varphi^{ \prime }
+
\cos \varphi^{ \prime } \cdot \partial_{ r^{ \prime } r^{ \prime } }
+
\frac{ \sin \varphi^{ \prime } }{ { r^{ \prime } }^2 }
\cdot
\partial_{ \varphi^{ \prime } }
-
\frac{ \sin \varphi^{ \prime } }{ r^{ \prime } } \cdot
\partial_{ r^{ \prime } \varphi^{ \prime } }
+
i k \cos \theta_{ t m }
\cos \varphi^{ \prime } \cdot \partial_{ z^{ \prime } } \\
& \partial_{ z^{ \prime } r^{ \prime } }
- i k \cos \theta_{ t m } \cdot \partial_{ r^{ \prime } }  \\
\end{aligned}
\right\}
G( \tilde{ R } ),
\end{aligned}
\label{eq:(30)}
\end{equation}
\end{widetext}
where
$ m $
is a number of considered wave,
$ \theta_{ t m } $
is the corresponding angle of refraction.
Note that, obtaining the result of~%
\eqref{eq:(30)}%
, we used the properties of the evenness and oddness of various terms in the integrands (in particular, this leads to zeroing
$ E_{ \varphi }^{ ( m ) } $%
).
The total radiation field is the sum of these components:
$ \vec{ E } \approx \sum\nolimits_m \vec{ E }^{ ( m ) } $.

The integration limits
$ r_{ m l } $,
$ r_{ m h } $
are determined by the limits of the aperture that is the cone base part which is illuminated by the wave under consideration.
For two waves under consideration
\begin{equation}
\begin{aligned}
r_{ 1 l } &= a, \\
r_{ 2 l } &= \max ( a, a + l \tan \theta_i ), \\
r_{ 1 h } & = r_{ 2 h } = a + b.
\end{aligned}
\label{eq:(31)}
\end{equation}
The formula for
$ r_{ 2 l } $
is explained by the fact that the illuminated part of the cone base is smaller than the entire base in the case of
$ \theta_{ i 2 } > 0 $.

Further one can exactly find all derivatives in~%
\eqref{eq:(30)}%
, but the result will be very cumbersome.
On the other hand, the exact calculation is not very important, because, as a rule, we are interested in the field on the distance much larger than wavelength under consideration.
Assuming that
$ k \left| z - l_0 \right| \gg 1 $
and, therefore,
$ k \tilde{R} \gg 1 $
for all values of
$ r^{ \prime }, \varphi^{ \prime } $,
we can differentiate only
$ \exp \left( i k \tilde{ R } \right) $
in the function
$ G \left( \tilde{ R } \right) $.
As a result, the formulas~%
\eqref{eq:(30)}
are reduced to the following one:

%
%%%%%%%%%%%%%%%%%%%%%%%%%%%%%%%%%%%%%%%%%%%%%%%%%%%%%%%%%%%%%%%%%%%%%%%%%%%%%%%%%%%%%%%%%
\begin{figure}[t]
\centering
\includegraphics[width=0.85\linewidth]{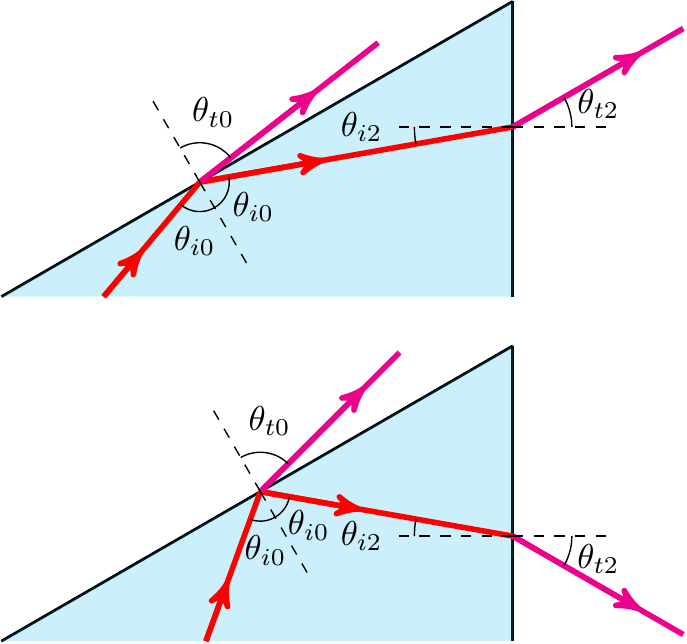}
\caption{\label{fig:2}%
The path of the ray for the case
${{\theta }_{i}}>0$
(top) and
${{\theta }_{i}}<0$
(bottom).
}
\end{figure}
%%%%%%%%%%%%%%%%%%%%%%%%%%%%%%%%%%%%%%%%%%%%%%%%%%%%%%%%%%%%%%%%%%%%%%%%%%%%%%%%%%%%%%%%%%%%
%

\begin{widetext}
\begin{equation}
\left\{
\begin{aligned}
& E_r^{ ( m ) } \\
& E_z^{ ( m ) } \\
\end{aligned}
\right\}
=
-\frac{ i k }{ 2 \pi }
\int\limits_{ r_{ m l } }^{ r_{ m h } }
dr^{ \prime }
\int\limits_0^{ \pi }
d \varphi^{ \prime } r^{ \prime }
H_{ \varphi^{ \prime } }^{ ( m ) }( \vec{ R }^{ \prime } )
\frac{ e^{ i k \tilde{ R } } }{ \tilde{ R }^3 }
\left\{
\begin{aligned}
& ( z - l_0 )^2 \cos \varphi^{ \prime } + r ( r^{ \prime } - r \cos \varphi^{ \prime } )
\sin^2 \varphi^{ \prime } + ( z - l_0 ) \tilde{ R }
\cos \theta_{ t m } \cos \varphi^{ \prime } \\
& ( r^{ \prime } - r \cos \varphi^{ \prime } )
( \tilde{ R } \cos \theta_{ t m } + z - l_0 ) \\
\end{aligned}
\right\}.
\label{eq:(32)}
\end{equation}
Using the expressions~%
\eqref{eq:(10)}
and
\eqref{eq:(22)}
for
$ H_{ \varphi^{ \prime } }^{ ( m ) } $
we obtain
\begin{equation}
\left\{
\begin{aligned}
& E_r^{ ( m ) } \\
& E_z^{ ( m ) } \\
\end{aligned}
\right\}
=
-\frac{ i Q_m }{ 2 \pi }
\int\limits_{ r_{ m l } }^{ r_{ m h } }
d r^{ \prime }
\int\limits_0^{ \pi } d \varphi^{ \prime }
\frac{ \sqrt{ k r^{ \prime } }
e^{ i \Phi_m ( r^{ \prime }, \varphi^{ \prime } ) } }
{ \tilde{ R }^3 }
\left\{
\begin{aligned}
& ( z - l_0 )^2 \cos \varphi^{ \prime } + r ( r^{ \prime } - r \cos \varphi^{ \prime } )
\sin^2 \varphi^{ \prime } + ( z - l_0 )
\tilde{ R } \cos \theta_{ t m } \cos \varphi^{ \prime } \\
& ( r^{ \prime } - r \cos \varphi^{ \prime } )
( \tilde{ R } \cos \theta_{ t m } + z - l_0 ) \\
\end{aligned}
\right\},
\label{eq:(33)}
\end{equation}
\end{widetext}
where
\begin{equation}
\Phi_m ( r^{ \prime }, \varphi^{ \prime } ) =
k r^{ \prime } \sin \theta_{ t m }
+
k \tilde{ R } ( r^{ \prime }, \varphi^{ \prime } ).
\label{eq:(34)}
\end{equation}

%
%
%%%%%%%%%%%%%%%%%%%%%%%%%%%%%%%%%%%%%%%%%%%%%%%%%%%%%%%%%%%%%%%%%%%%%%%%%%%%%%%%%%%%%%%%
%%%%%%%%%%%%%%%%%%%%%%%%%%%%%%%%%%%%%%%%%%%%%%%%%%%%%%%%%%%%%%%%%%%%%%%%%%%%%%%%%%%%%%%%
\section{\label{sec:rayoptics} Ray optics approximation }
%%%%%%%%%%%%%%%%%%%%%%%%%%%%%%%%%%%%%%%%%%%%%%%%%%%%%%%%%%%%%%%%%%%%%%%%%%%%%%%%%%%%%%%%
%%%%%%%%%%%%%%%%%%%%%%%%%%%%%%%%%%%%%%%%%%%%%%%%%%%%%%%%%%%%%%%%%%%%%%%%%%%%%%%%%%%%%%%%

%
%%%%%%%%%%%%%%%%%%%%%%%%%%%%%%%%%%%%%%%%%%%%%%%%%%%%%%%%%%%%%%%%%%%%%%%%%%%%%%%%%%%%%%%%%
\begin{figure*}[t]
\centering
\includegraphics[width=0.85\linewidth]{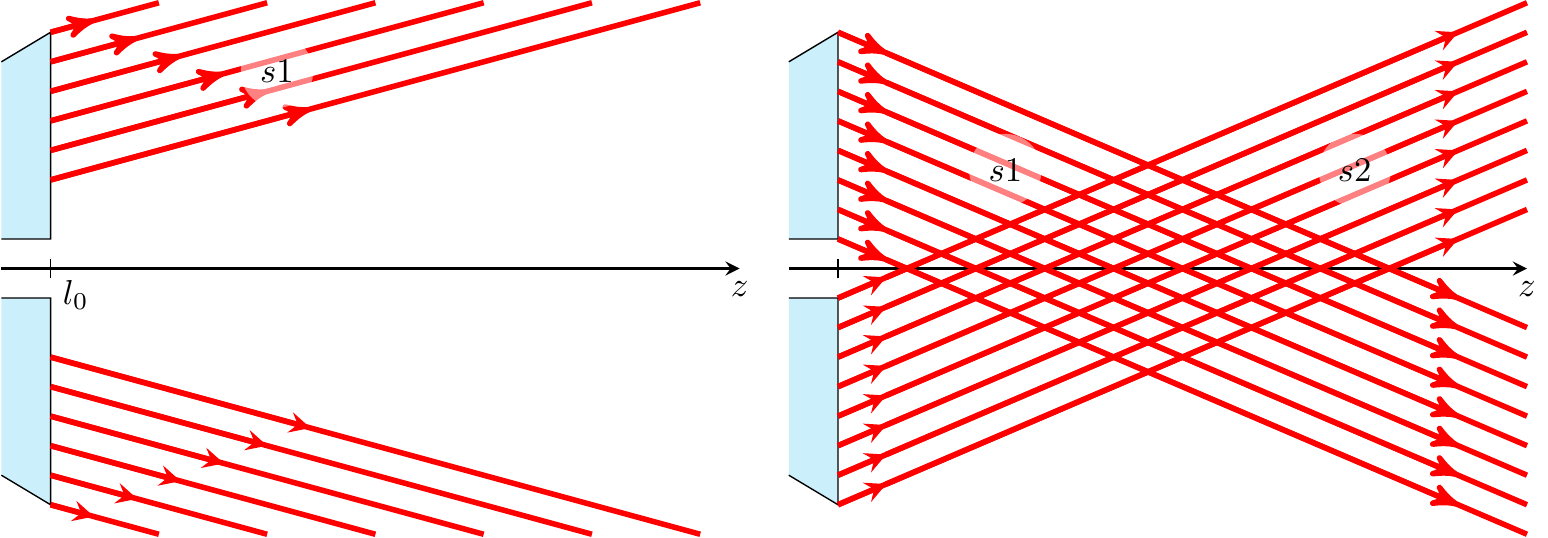}
\caption{\label{fig:3}%
The ray picture outside the cone in the cases of
${{\theta }_{i2}}>0$
(left) and
${{\theta }_{i2}}<0$ (right).
}
\end{figure*}
%%%%%%%%%%%%%%%%%%%%%%%%%%%%%%%%%%%%%%%%%%%%%%%%%%%%%%%%%%%%%%%%%%%%%%%%%%%%%%%%%%%%%%%%%%%%

Let us find the saddle point (or stationary phase point) for the integrands in~%
\eqref{eq:(33)}%
.
This point is determined by equations~%
\cite{FMb}
\begin{equation}
\frac{\partial \Phi_m ( r^{ \prime }, \varphi^{ \prime } ) }
{ \partial r^{ \prime } }
= 0,
\quad
\frac{ \partial \Phi_m ( r^{ \prime }, \varphi^{ \prime } ) }
{\partial \varphi^{ \prime } } = 0.
\label{eq:(35)}
\end{equation}
It is easily to find that this system has the following two solutions:
\begin{equation}
\begin{aligned}
& r^{ \prime }
= r_m^{ s 1 }
= r - ( z - l_0 ) \tan \theta_{ t m },
\quad
\varphi^{ \prime }
= \varphi_m^{ s 1 } = 0; \\
& r^{ \prime }
= r_m^{ s 2 }
= - r - ( z - l_0 ) \tan \theta_{ t m },
\quad
\varphi^{ \prime } = \varphi_m^{ s 2 } = \pi.
\end{aligned}
\label{eq:(36)}
\end{equation}
Since
$ \theta_{ t 1 } > 0 $,
then
$ r_1^{ s 2 } < 0 $,
and this saddle point lies beyond the integration limits.
Therefore the first wave is determined only by the saddle point
$ s 1 $
with
$ r^{ \prime } = r_1^{ s 1 } $.
At the same time, the value
$ \theta_{ t 2 } $
can be both positive and negative.
Therefore both saddle points
$ s1,2 $
can be significant for the second wave. 
First of all, we consider this wave.

Simple transformations give the following expressions for
$ \tilde{ R } $
and
$ \Phi_2 $
in the saddle points:
\begin{equation}
\begin{aligned}
&\tilde{ R } ( r_2^{ s 1 }, \varphi_2^{ s 1 } )
=
\tilde{ R } ( r_2^{ s 2 }, \varphi_2^{ s 2 } )
=
( z - l_0 ) / \cos \theta_{ t 2 }, \\
&\Phi_2^{ s 1, 2 }
=
\Phi_2 ( r_2^{ s 1, 2 }, \varphi_2^{ s 1, 2 } )
=
k
\left[
\pm r \sin \theta_{ t 2 } + ( z - l_0 ) \cos \theta_{ t 2 }
\right].
\end{aligned}
\label{eq:(37)}
\end{equation}
Further we will need as well values of the second derivatives of the phase in the saddle points:
\begin{equation}
\begin{aligned}
\left.
\frac{ \partial^2 \Phi_2 }
{ \partial { r^{ \prime } }^2 }
\right|_{ s 1, 2 }
&=
k
\frac{ \cos^3 \theta_{ t 2 } }{ z - l_0 },
\quad
\left.
\frac{ \partial^2 \Phi_2 }
{ \partial { \varphi^{ \prime } }^2 }
\right|_{ s 1, 2 }
= \\
&=
\pm
\frac{ r r_2^{ s 1, 2 } \cos \theta_{ t 2 } }
{ z - l_0 },
\quad
\left.
\frac{ \partial^2 \Phi_2 }
{ \partial \varphi^{ \prime } \partial r^{ \prime } }
\right|_{ s 1, 2 }
= 0.
\end{aligned}
\label{eq:(38)}
\end{equation}

We can approximately calculate the integrals~%
\eqref{eq:(33)}
by the stationary phase method if the aperture contains a large number of Fresnel zones, in other words, the function
$ e^{ i \Phi_2 ( r^{ \prime }, \varphi^{ \prime } ) } $
experiences a large number of oscillations within this area. 
This condition means that
$ \left|
\Phi_2 ( r^{ \prime }, \varphi^{ \prime } ) - \Phi_2^{ s 1, 2 }
\right| \gg 1 $
on the most part of the aperture.
We can write this inequality as
$ \left|
\frac{ \partial^2 \Phi_2 }
{\partial { r^{ \prime } }^2 }{ b^2 }
\right| \gg 1$.
If
$  \cos \theta_{ t 2 } $
is not very small, then we obtain
$ \left|
\frac{ k b^2 }{ z - l_0 }
\right| \gg 1 $,
or
\begin{equation}
D \sim \frac{ \lambda ( z - l_0 ) }{ \pi b^2 } \ll 1.
\label{eq:(39)}
\end{equation}
%
%where
%$ \lambda = 2 \pi / k $
%is a wavelength.
The parameter
$D$ is usually called a ``wave parameter''~%
\cite{KravOrb}.
The inequality~%
\eqref{eq:(39)}
is the condition of applicability of the ray optics approximation.
	
Applying the known expression for asymptotic of double integral~%
\cite{FMb}
one can obtain the following result:

\ 

\begin{widetext}
\begin{equation}
\begin{aligned}
&\vec{ E }^{ ( 2 ) }
\approx
\vec{ E }^{ ( s 1 ) } + \vec{ E }^{ ( s 2 ) } \\
&\left\{
\begin{aligned}
& E_r^{ s 1 } \\
& E_z^{ s 1 } \\
\end{aligned}
\right\}
=
Q_2
\frac{ e^{ i \Phi_2^{ s 1 } } }{ \sqrt{ k r } }
\left\{
\begin{aligned}
&\cos \theta_{ t 2 } \\
&-\sin \theta_{ t 2 } \\
\end{aligned}
\right\}
\left\{
\begin{aligned}
& 1  \text{ for }  r_{ 2 l } < r - ( z - l_0 ) \tan \theta_{ t 2 } < r_{ 2 h }, \\
& 0  \text{ otherwise }
\end{aligned}
\right\}, \\
&\left\{
\begin{aligned}
& E_r^{ s 2 } \\
& E_z^{ s 2 } \\
\end{aligned}
\right\}
=
Q_2
\frac{ e^{ i \Phi_2^{ s 2 } } }{ \sqrt{ k r } }
\left\{
\begin{aligned}
& \cos \theta_{ t 2 } \\
& \sin \theta_{ t 2 } \\
\end{aligned}
\right\}
\left\{
\begin{aligned}
& 1  \text{ for }  -r_{ 2h } < r + ( z - l_0 ) \tan \theta_{ t 2 } < -r_{ 2l }, \\
& 0  \text{ otherwise }
\end{aligned}
\right\}, \\
\end{aligned}
\label{eq:(40)}
\end{equation}
where
\begin{equation}
\Phi_2^{ s 1 }
=
k r \sin \theta_{ t 2 }
+
k ( z - l_0 ) \cos \theta_{ t 2 },
\quad
\Phi_2^{ s 2 }
=
- k r \sin \theta_{ t 2 }
+
k ( z - l_0 )
\cos \theta_{ t 2 } - \pi / 2.
\label{eq:(41)}
\end{equation}
\end{widetext}
One can see that the contributions of stationary points exist only in certain regions shown in Fig.~%
\ref{fig:3}
(their borders are ray optics boundaries).
These limitations are explained by the fact that that only under such conditions the stationary phase points are in the limits of integration (on the aperture), i. e.
$ r_{ 2 l } < r_2^{ s 1, 2 } < r_{ 2 h } $.
If this condition is violated for one of the stationary points, then this point is outside the aperture, and its contribution is zero.
More precisely one can say that the ray optics solution~%
\eqref{eq:(40)}
is suitable at some distance from the ray optics boundaries exceeding the wavelength.
	
The Eq.~%
\eqref{eq:(40)} 
describes two quasi-plane waves (more precisely, they are cylindrical waves with small curvature of the constant phase surface). Naturally, these waves are transverse because the projections on the propagation direction are zero:
$ E_{ \parallel }^{ s 1, 2 } =
\pm E_{r}^{ s 1, 2 } \sin \theta_{ t 2 }
+
E_{ z }^{ s 1, 2 } \cos \theta_{ t 2 } = 0 $.
The electric field is orthogonal to the propagation direction:
\begin{equation}
E_{ \bot }^{ s 1, 2 } = H_{ \varphi }^{ s 1, 2 } 
= 
Q_2 \exp \left( i \Phi_2^{ s 1, 2 } \right) \left/ \sqrt{ k r } \right..
\label{eq:(42)}
\end{equation}
The wave ``s1'' exists for any sign of the angles 
$ \theta_{ i 2} $,  
$\theta_{ t 2 } $ 
and propagates at the angle
$ \theta_{ t 2 } $ 
with respect to the 
$z$%
-axis (Fig.~%
\ref{fig:3}, 
left and right).
The wave ``s2'' exists only in the case 
$ \theta_{ i 2 }, \theta_{ t 2 } < 0 $ 
and
propagates at the angle 
$ \left| \theta_{ t 2 } \right| = - \theta_{ t 2 } $ 
with respect to the
$z$%
-axis (Fig.~%
\ref{fig:3}, 
right).
Note that in the case
$ \theta_{ i 2 }, \theta_{ t 2 } < 0 $
(that is
$ 2 \alpha < \theta_p $%
), the rays converge to the
$ z $-axis, and there is certain rhomboidal area where the rays are intersected (Fig.~%
\ref{fig:3}%
, right).
In this area, the ray optics solution~%
\eqref{eq:(40)}
tends to infinity if
$ r \to 0 $
on the segment
$ r_{ 2 l } / \tan \left| \theta_{ t 2 } \right|
<
z - l_0
<
r_{ 2 h } / \tan \left| \theta_{ t 2 } \right| $.
This means that the ray optics approximation is not applicable at distances from the
$ z $%
-axis less than the wavelength under consideration.
However we can expect that the real field has a large value in this area.
	
Naturally, the expressions~%
\eqref{eq:(40)}
can be obtained with help of the ray-optics method.
Let us give this derivation briefly.
The wave exiting the target is a quasi-plane transversal wave having the electric and magnetic fields
egual each other and determined by the formula~%
\eqref{eq:(22)}
on the aperture.
Because of axial symmetry the exiting wave is cylindrical.
Considering also that the boundary of the object in its section is a straight line, it is easy to conclude that the wave amplitude in the point
$ ( r, z ) $
differs from one in the point
$  ( r^{ \prime }, l_0 )$
only by replacement
$ r^{ \prime } $
to
$ r $
(similar effect is discussed in~%
\cite{BTG13}
for other objects).
Thus the formula~%
\eqref{eq:(22)}
results in the expression
$ \left| E^{(2)} \right|
=
\left| Q_2 \right| / \sqrt{ k r } $
which corresponds to Eq.~%
\eqref{eq:(40)}.

It remains to determine the phases of two waves.
First we consider the wave ``s1'' radiated from the upper part of the aperture.
Taking into account that the length of the ray outside the target is
$ ( z - l_0 ) / \cos \theta_{ t 2 } $
we have for the phase at the point
$ ( r, z ) $
\begin{equation}
\Phi_2 ( r, z )
=
\Phi_{ i 2 } ( r^{ \prime }, l_0 )
+
k \frac{ z - l_0 }{ \cos \theta_{ t 2 } },
\label{eq:(43)}
\end{equation}
where
$ \Phi_{ i 2 }( r^{ \prime }, l_0 ) $
is given by the formula~%
\eqref{eq:(20)}
with
$ r^{ \prime } = r - ( z - l_0 ) \tan \theta_{ t 2 } $.
Substituting~%
\eqref{eq:(20)}
in
\eqref{eq:(43)}
one can obtain that
\begin{equation}
\begin{aligned}
\Phi_2 ( r, z )
&=
k
\left[
n l_0 \cos \theta_{ i 2 }
+
r \sin \theta_{ t 2 }
+
( z - l_0 ) \cos \theta_{ t 2 }
\right] - \\
&-
\pi / 4
=
\arg Q_2 + \Phi_2^{ s 1 },
\end{aligned}
\label{eq:(44)}
\end{equation}
which corresponds to~%
\eqref{eq:(40)},
\eqref{eq:(41)}.
	
Similar way gives corresponding result for the wave ``s2'' if we take into account that for this wave
$ \theta_{ t 2 } < 0 $.
However, we should take into account the following difference.
The ray ``s2'' passes through the
$ z $%
-axis, which is a caustic where the ray tube cross-section tends to zero.
It is known~%
\cite{KravOrb}
that during the passage of the caustic, the phase of the wave changes to
$ \pi / 2 $.
Taking into account this factor, we obtain
$ \Phi_2 ( r, z ) = \arg Q_2 + \Phi_2^{ s 2 } $,
where
$ \Phi_2^{ s 2 } $
is given by Eq.~%
\eqref{eq:(41)}.

Until now in this section, we have considered only the second wave (that is, the wave reflected from the lateral wall).
The ray-optical analysis of the first wave is simpler, since it is determined by one saddle point ``s1'' only.
By analogy with formulas~%
\eqref{eq:(40)}%
, we obtain
\begin{equation}
\begin{aligned}
&\vec{ E }^{ ( 1 ) }
\approx
\vec{ E }^{ s 1 } + \vec{ E }^{ s 2 }, \\
&\left\{
\begin{aligned}
&E_r^{ s 1 } \\
&E_z^{ s 1 }
\end{aligned}
\right\}
=
Q_1
\frac{ e^{ i \Phi_1^{ s 1 } } }{ \sqrt{ k r } }
\left\{
\begin{aligned}
&\cos \theta_{ t 1 } \\
&- \sin \theta_{ t 1 }
\end{aligned}
\right\}
\times \\
&\times
\left\{
\begin{aligned}
&1
\,\,\,
\text{ for }
\,\,\,
r_{ 1 l }< r - ( z - l_0 ) \tan \theta_{ t 1 } < r_{ 1 h } \\
&0
\,\,\,
\text{ otherwise }
\end{aligned}
\right\},
\end{aligned}
\label{eq:(45)}
\end{equation}
where
\begin{equation}
\Phi_1^{ s 1 }
=
k r \sin \theta_{ t 1 }
+
k ( z - l_0 ) \cos \theta_{ t 1 }.
\label{eq:(46)}
\end{equation}

%%%%%%%%%%%%%%%%%%%%%%%%%%%%%%%%%%%%%%%%%%%%%%%%%%%%%%%%%%%%%%%%%
%%%%%%%%%%%%%%%%%%%%%%%%%%%%%%%%%%%%%%%%%%%%%%%%%%%%%%%%%%%%%%%%%
\section{\label{sec:fraunhofer} Fraunhofer area }
%%%%%%%%%%%%%%%%%%%%%%%%%%%%%%%%%%%%%%%%%%%%%%%%%%%%%%%%%%%%%%%%%
%%%%%%%%%%%%%%%%%%%%%%%%%%%%%%%%%%%%%%%%%%%%%%%%%%%%%%%%%%%%%%%%%

Now we consider the area where the wave parameter is large:
$ D \gg 1 $.
Usually this area is named Fraunhofer, or far-field, area.
Corresponding asymptotic can be obtained from both general approximate formulas~%
\eqref{eq:(5)}
and the expressions~%
\eqref{eq:(33)}
obtained for the geometry under consideration.

Based on Eq.~%
\eqref{eq:(33)}%
, we can use the approximation
$
\tilde{ R }
\approx
R_0
\left(
1 - r r^{ \prime } \cos \varphi^{ \prime } R_0^{ - 2 }
\right)
$
(here
$ R_0 = \sqrt{ r^2 + ( z - l_0 )^2 } $%
) in the phase
$ \Phi_m ( r^{ \prime }, \varphi^{ \prime } ) $
and rougher approximation
$ R \approx R_0 $
in other factors in the integrand:
\begin{widetext}
\begin{equation}
\begin{aligned}
&\left\{
\begin{aligned}
&E_r^{ ( m ) } \\
&E_z^{ ( m ) }
\end{aligned}
\right\}
=
-\frac{ i Q_m }{ 2 \pi }
\frac{ e^{ i k R_0 } }{ R^3 }
\int\limits_{ r_{ m l } }^{ r_{ m h } }
d r^{ \prime }
\int\limits_0^{ \pi }
d \varphi^{ \prime } \sqrt{ k r^{ \prime } } \times  \\
&\times
\left\{
\begin{aligned}
& z^2 \cos \varphi^{ \prime } - r^2 \cos \varphi^{ \prime }
\sin^2 \varphi^{ \prime }
+ z R \cos \theta_{ t m } \cos \varphi^{ \prime } \\
& -r ( R \cos \theta_{ t m } + z ) \cos \varphi^{ \prime } \\
\end{aligned}
\right\}
\exp \left( -i k \frac{ r }{ R_0 } r^{ \prime }
\cos \varphi^{ \prime }
+
i k r^{ \prime } \sin \theta_{ t m } \right),
\end{aligned}
\label{eq:(47)}
\end{equation}
Further it is convenient to use spherical coordinates
$ R $,
$ \theta $,
$ \varphi $.
Using the formulas
\begin{equation}
E_R = E_r \sin \theta + E_z \cos \theta,
\quad
E_{ \theta } = E_r \cos \theta - E_z \sin \theta
\label{eq:(48)}
\end{equation}
one can obtain
\begin{equation}\label{eq:(49)}
\left\{
\begin{aligned}
&E_R^{(m)} \\
&E_{ \theta }^{(m)}
\end{aligned}
\right\}
=
-\frac{ i Q_m }{ 2 \pi }
\frac{ e^{ i k R_0 } }{ R }
\int\limits_{ r_{ m l } }^{ r_{ m h } }
d r^{ \prime } \sqrt{ k r^{ \prime } }
\left\{
\begin{aligned}
& \left[ I_3 \left( \chi \right) - I_1 \left( \chi \right) \right]
\sin^3 \theta  \\
& I_1 \left( \chi \right)( \cos \theta_{ t m } + \cos^3 \theta )
+
I_3 \left( \chi \right) \sin^2 \theta \cos \theta
\end{aligned}
\right\}
e^{ i k r^{ \prime } \sin \theta_{ t m } },
\end{equation}
\end{widetext}
where
\[
I_m \left( \chi \right)
=
\int\nolimits_0^{ \pi }
e^{ - i \chi \cos x } \cos^m x d x
\]
and
$ \chi = k r^{ \prime } r R_0^{ -1 }
\approx k r^{ \prime } \sin \theta $.
The integrals
$ I_m ( \chi ) $
are known, they are expressed in terms of Bessel functions
$ J_m ( \chi ) $~%
\cite{PBMb1}:
\begin{equation}
\begin{aligned}
I_1 ( \chi )
&=
- \pi i J_1 ( \chi ), \\
I_3 ( \chi )
&=
\pi i
\left[
\left( \frac{ 2 }{ \chi^2 } - 1 \right)
J_1 ( \chi ) - \frac{ J_0( \chi ) }{ \chi }
\right].
\end{aligned}
\label{eq:(50)}
\end{equation}
Asymptotes of the functions~%
\eqref{eq:(50)}
for
$ \chi \gg 1 $
are the same:
\begin{equation}
I_1 ( \chi )
\approx
I_3 (\chi )
\approx
-i \sqrt{ \frac{ 2 \pi }{ \chi } }
\cos \left( \chi -\frac{ 3 \pi }{ 4 } \right),
\label{eq:(51)}
\end{equation}
and the error has the order of
$ O \left( \chi^{ -3 / 2 } \right)$.
One can see that
\begin{equation}
\left[
I_3 ( \chi ) -
I_1 ( \chi )
\right]
\sin^3 \theta
{=}
O
\left(
\frac{ \sin^3 \theta }{ \chi^{ 3 / 2 } }
\right)
{=}
O
\left(
\frac{ \sin^{ 3 / 2 } \theta }{ ( k r^{ \prime } )^{ 3 / 2 } } \right).
\label{eq:(52)}
\end{equation}
Since
$ k r^{ \prime } \sim k b \gg 1 $
on the almost all aperture then we obtain that
$ \left| E_R \right| \ll \left| E_{ \theta } \right| $,
and
$ R E_R \to 0 $
if
$ k b \to \infty $.
Thus the wave is practically transversal (that is natural), and further we consider the
$ \theta $%
-component only.

If the condition
$ k b \theta \gg 1 $
is true then
$ k r^{ \prime } \sin \theta \gg 1 $
on the almost whole aperture, and using~%
\eqref{eq:(51)}
we obtain from~%
\eqref{eq:(49)}
the following result:
\begin{equation}
E_{ \theta }^{ ( m ) }
\approx
H_{ \varphi }^{ ( m ) }
\approx
-\frac{ Q_m }{ \sqrt{ 2 \pi } }
\frac{ \cos \theta + \cos \theta_{ t m } }
{ \sqrt{ \sin \theta } }
F_m ( \theta )
\frac{ d_m }{ R }
e^{ i k R_0 },
\label{eq:(53)}
\end{equation}
where
\begin{equation}
\begin{aligned}
&F_m( \theta )
{=}
\frac{ 1 }{ d_m }
\int\limits_{ r_{ m l } }^{ r_{ m h } }
\cos
\left(
k r^{ \prime } \sin \theta
{-}
\frac{ 3 \pi }{ 4 }
\right)
\exp \left( i k r^{ \prime } \sin \theta_{ t m } \right)
d r^{ \prime }
{=} \\
&=
\frac{ \sin ( d_m w_{ m - } ) }
{ d_m w_{ m - } }
e^{ i \bar{ r }_m w_{ m - } + 3 i \pi / 4 }
+
\frac{ \sin ( d_m w_{ m + } ) }
{ d_m w_{ m + } }
e^{ i \bar{ r }_m w_{ m + } - 3 i \pi / 4 }
\end{aligned}
\label{eq:(54)}
\end{equation}
\begin{equation}
\begin{aligned}
w_{ m \pm }
&=
k \left( \sin \theta_{ t m } \pm \sin \theta \right), \\
d_m
&=
\frac{ r_{ m h } {-} r_{ m l } }{ 2 },
\quad
\bar{ r }_m = \frac{ r_{ m h } {+} r_{ m l } }{ 2 }.
\end{aligned}
\label{eq:(55)}
\end{equation}
\[
\begin{aligned}
d_1 &= b / 2,
\quad
\bar{ r }_1 = b / 2 + a, \\
d_2
&\approx
\left\{
\begin{aligned}
&\left(
b - l \tan \theta_{ i 2 }
\right)/2
\quad \text{ for } \quad \theta_{ i 2 } {>} 0, \\
& b / 2
\quad \text{ for } \quad \theta_{ i 2 } {<} 0,
\end{aligned}
\right., \\
\bar{ r }_2
&=
\left\{
\begin{aligned}
&a + \left( b + l \tan \theta_{ i 2 } \right) / 2
\quad \text{ for } \quad \theta_{ i 2 } {>} 0, \\
&a + b / 2
\quad \text{ for } \quad \theta_{ i 2 } {<} 0.
\end{aligned}
\right.
\end{aligned}
\]
%
%Underline that this result is true if
%$ k b \theta \gg 1 $.
%Naturally, we obtained the spherical transversal wave where
%${{E}_{\theta }}={{H}_{\varphi }}$.	
	
The radiation pattern is determined primarily by the function
$ F_m ( \theta ) $.
Since
$ \theta_{ t 1 } > 0 $
then the function
$ F_1 ( \theta ) $
has the main maximum at
$ \theta = \theta_{ t 1 } $
(in fact, only the first summand in~%
\eqref{eq:(54)}
has the importance for the function
$ F_1 ( \theta ) $%
).

The behavior of the function
$ F_2 ( \theta ) $
is more complex.
If
$ \theta_{ t 2 } > 0 $
then the main maximum of the function
$ F_2 ( \theta ) $
is determined by the first summand in~%
\eqref{eq:(54)}%
: it takes place for
$ \theta = \theta_{ t 2 } $
(radiation comes mainly from the ``upper'' part of the aperture, as it is shown in the left plot of Fig.~%
\ref{fig:2}%
).
If
$ \theta_{ t 2 } < 0 $
then the main maximum of the function
$ F_2 ( \theta ) $
is determined by the second summand in~%
\eqref{eq:(54)}%
: it takes place for
$ \theta = - \theta_{ t 2 } = \left| \theta_{ t 2 } \right| $
(radiation comes mainly from the ``lower'' part of the aperture, as it is shown in the right plot of Fig.~%
\ref{fig:2}%
).

Thus, the direction of maximal radiation coincides with the direction of refraction wave (this is natural).
In any case, the maximum values of
$\left| F_m ( \theta ) \right|$
is approximately equal to
$ 1 $%
, and maxima of the fields are equal to
\begin{equation}
\left|
E_{ \theta }^{ ( m ) }
\right|_{ \max }
\approx
\frac{ \left| Q_m \right| }{ \sqrt{ 2 \pi } }
\frac{ 2 \cos \theta_{ t m } }{ \sqrt{ \sin \theta_{ t m } } }\frac{ d_m }{ R }.
\label{eq:(56)}
\end{equation}
The angular width
$ \delta \theta $
of the main lobes of the diagrams is
$ \delta \theta \approx \frac{ 2 \pi }{ k d \cos \theta_{ t m } }$.

%%%%%%%%%%%%%%%%%%%%%%%%%%%%%%%%%%%%%%%%%%%%%%%%%%%%%%%%%%%%%%%%%
%%%%%%%%%%%%%%%%%%%%%%%%%%%%%%%%%%%%%%%%%%%%%%%%%%%%%%%%%%%%%%%%%
\section{\label{sec:spotlight} ``Cherenkov spotlight'' }
%%%%%%%%%%%%%%%%%%%%%%%%%%%%%%%%%%%%%%%%%%%%%%%%%%%%%%%%%%%%%%%%%
%%%%%%%%%%%%%%%%%%%%%%%%%%%%%%%%%%%%%%%%%%%%%%%%%%%%%%%%%%%%%%%%%

\begin{figure}
  \includegraphics{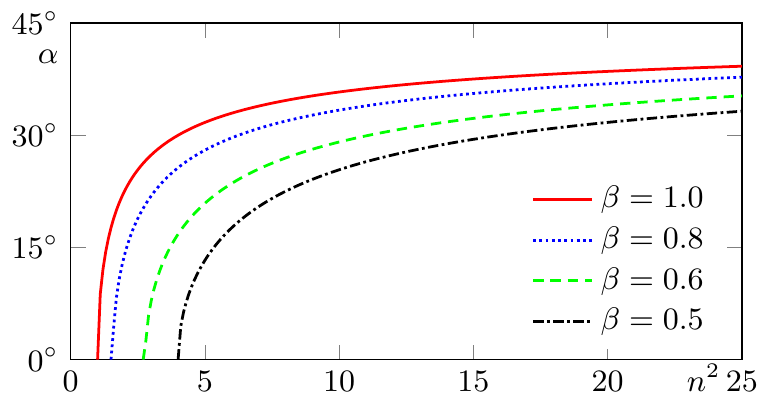}%
  \caption{\label{fig:4} The cone angle for the spotlight effect depending on the refractive index square $n^2=\varepsilon \mu$ for different value of the  charge velocity.}
\end{figure}

Note that the expressions~%
\eqref{eq:(53)}%
,
\eqref{eq:(53)}
are not true for
$ k b \theta \le 1 $.
However this angle range is very interesting in the important case when the second wave propagates along the symmetry axis that is
$ \theta_{ i 2 } = \theta_{ t 2 } = 0 $
(for the first wave this situation is impossible).
According to~%
\eqref{eq:(16)}%
, this situation takes place when
\begin{equation}
\alpha = \frac{ \theta_p }{ 2 }
= \left.
\mathrm{arccos} \left( ( n \beta )^{-1} \right)
\right/ 2 .
\label{eq:(57)}
\end{equation}
Figure~%
\ref{fig:4} 
shows dependency of the cone angle~%
\eqref{eq:(57)} 
on the refractive index for different value of the charge velocity.
Let us consider this case separately, assuming, as in the previous section, that the observation point is in the Fraunhofer region (%
$ D \gg 1 $%
).

The integrand in~%
\eqref{eq:(49)}
contains Bessel functions, which can be represented in the form of Taylor series~%
\cite{ASb}.
After that, in the case
$ \theta_{ i 2 } = \theta_{ t 2 } = 0 $
we obtain
\begin{equation}
E_{ \theta }^{ ( 2 ) }
=
-\frac{ i Q_2 }{ 2 \pi }
\frac{ e^{ i k R_0 } }{ R }
\int\limits_{ r_{ 2 l } }^{ r_{ 2 h } }
U( r^{ \prime } ) d r^{ \prime },
\label{eq:(58)}
\end{equation}
where
$ r_{ 2 l } = a $,
$ r_{ 2 h } = a + b \approx b $,
\begin{equation}
\begin{aligned}
U( r^{ \prime } )
&=
\pi i
\sum\limits_{ m = 0 }^{ \infty }
\frac{ ( -1 )^{m + 1 }
\left( \sin \theta \right)^{ 2 m + 1 } }
{ 2^{ 2 m } m! (
m + 1 ) ! }
( k r^{ \prime } )^{ 2 m - 1 / 2 } {\times} \\
&\times
\left[
m \cos \theta +
\left(
k r^{ \prime } \cos \frac{ \theta }{ 2 }
\right)^2
\right].
\end{aligned}
\label{eq:(59)}
\end{equation}
Calculating the integrals of the terms of the series and considering that
$ a \ll b $,
we obtain the following result:
\begin{equation}
E_{ \theta }^{ ( 2 ) }
=
-\frac{ Q_2 }{ 5 }
\left( k b \right)^{ 3 / 2 } F_0 ( \theta )
\frac{ e^{ i k R_0 } }{ k R },
\label{eq:(60)}
\end{equation}
where
\begin{equation}
F_0 ( \theta )
=
\sum\limits_{m = 0 }^{ \infty }
\frac{ ( -1 )^m  \left( k b \sin \theta  \right)^{ 2 m + 1 } }
{\left( 4 m / 5 + 1 \right) 2^{ 2 m } m ! ( m + 1 ) ! }
\cos^2 \frac{ \theta }{ 2 }.
\label{eq:(61)}
\end{equation}
The function
$ F_0 ( \theta ) $
determines the radiation pattern for the considered case where
$ \theta_{ i 2 } = { \theta }_{ t 2 } = 0 $.
Note that, under condition
$ k b \sin \theta \le 1 $
or
$ \theta \le ( k b )^{ -1 } \ll 1 $,
one can use the simple approximation
\begin{equation}
F_0 (\theta )
\approx
k b \theta
\left[ 1 - \frac{ 5 }{ 72 }
\left( k b \theta \right)^2
\right].
\label{eq:(62)}
\end{equation}
The angle of the maximum
$ \theta_{ \max } $
and the maximal value for this function are equal to
\begin{equation}
\theta_{ \max } \approx 2.19 / ( k b ),
\quad
F_{ 0 \max } = F_0 ( \theta_{ \max } ) \approx 1.46.
\label{eq:(63)}
\end{equation}

It is interesting to compare the maximal value of the field in the case when
$ \theta_{ i 2 } = 0 $
and in the case when
$ \theta_{ i 2 } \sim 1 $.
Based on~%
\eqref{eq:(60)},
\eqref{eq:(62)}
and
\eqref{eq:(56)}
we obtain
\begin{equation}
\begin{aligned}
\frac{
\left.
\left| E_{ \theta }^{ ( 2 ) } \right|_{\max}
\right|_{ \theta_{ i 2 } = 0 }
}
{
\left.
\left|
E_{ \theta }^{ ( 2 ) }
\right|_{\max}
\right|_{ \theta_{ i 2 } \sim 1 }
}
&\approx
\frac{ \sqrt{ 2 \pi } }{ 10 }
\frac{ \sqrt{ \sin \theta_{ t 2 } } }
{ \cos \theta_{ t 2 } }
F_{ 0 \max }
\frac{ ( k b )^{ 3 / 2 } }{ k d_2 }
{\sim} \\
&\sim
\sqrt{ k b } \gg 1.
\end{aligned}
\label{eq:(64)}
\end{equation}
Thus, if
$ \theta_{ i 2 } = 0 $,
then the field maximum is located at the small angle~%
\eqref{eq:(63)}%
, and its value is much larger than that for
$ \theta_{ i 2 } \sim 1 $.
Such an effect can be called ``Cherenkov spotlight''.

Note that the similar phenomenon occurs also for the case when the charge flies into the cone from the side of its base (``ordinary'' cone)~%
\cite{TGV19}.
However there is strong difference of conditions for reaching the effect.
In the ``ordinary'' case the ``spotlight effect'' is possible only in certain very narrow range of charge velocities close to the speed of light in the medium~%
\cite{TGV19}.
In the case under consideration (``inverted'' cone), this effect can be achieved for any charge velocity
$ \beta > 1 / n $
due to the proper selection of the cone angle
$ \alpha $
or refractive index
$ n $
in accordance with the condition~%
\eqref{eq:(57)}.

%%%%%%%%%%%%%%%%%%%%%%%%%%%%%%%%%%%%%%%%%%%%%%%%%%%%%%
%%%%%%%%%%%%%%%%%%%%%%%%%%%%%%%%%%%%%%%%%%%%%%%%%%%%%%
\section{\label{sec:numeric} Numerical results }
%%%%%%%%%%%%%%%%%%%%%%%%%%%%%%%%%%%%%%%%%%%%%%%%%%%%%%
%%%%%%%%%%%%%%%%%%%%%%%%%%%%%%%%%%%%%%%%%%%%%%%%%%%%%%

Here we demonstrate results of computation of the field in the most interesting far field (Fraunhofer) area. These results have been obtained with use of formula~%
\eqref{eq:(49)} 
which allows calculating the field everywhere in this region including the region of small angles 
$ \theta $ 
(therefore the case of the ``Cherenkov spotlight'' effect can be also analyzed in this way).

\begin{figure*}
\includegraphics{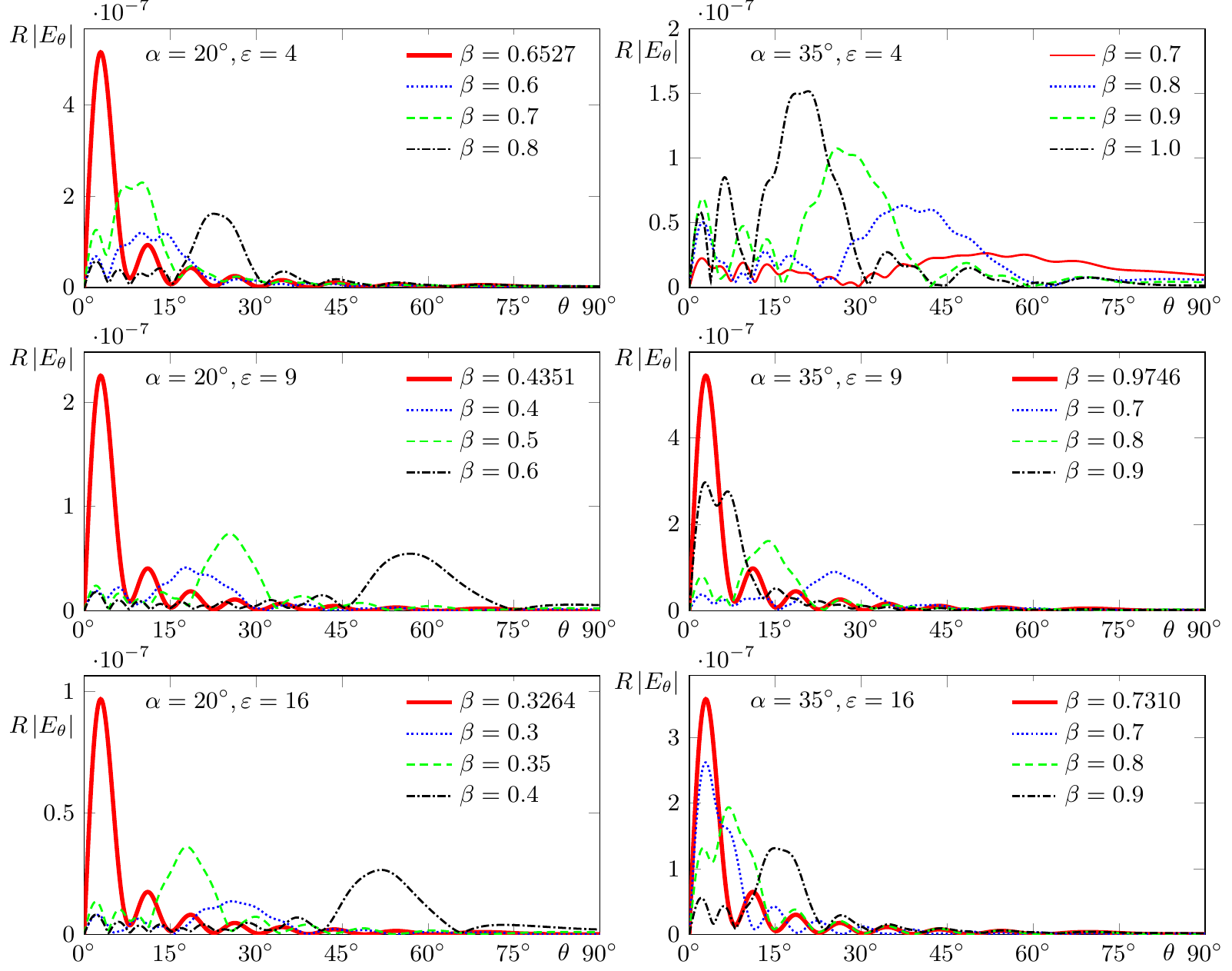}%
\caption{\label{fig:5} 
The angular distribution of the magnitude of the electric field Fourier-transform in the Fraunhofer area 
(in units 
$\mathrm{ V \cdot s }$%
). 
Parameters: $ a = c / \omega $, 
$ b = 50 c / \omega = 50 / k $, 
$ q = 1~\mathrm{nC} $, 
$ \mu=1 $; 
the cone angle 
$ \alpha $ 
and the charge velocity 
$ \beta $ 
are indicated in the plots.}
\end{figure*}

Figure~%
\ref{fig:5} 
shows the angle dependency of the field components for different values of the cone angle 
$ \alpha $ 
and the cone material permittivity 
$ \varepsilon $ 
(it is assumed that 
$ \mu = 1 $%
). 
The vertical axis on the plots shows the value 
$ R |E_\theta| $ 
which does not depend on the distance 
$ R $ 
in the Fraunhofer area.

Each plot contains four curves. 
For the bold red solid curve, the charge velocity corresponds to the condition~%
\eqref{eq:(57)} 
(i.e. 
$ \theta_{ t 2 } = 0 $%
) determining the ``Cherenkov spotlight'' effect. 
Other curves correspond to the cases when 
$ \theta_{ t 2 } \ne 0 $. 
One can see that the maximal field value is much larger for the ``spotlight'' case compared to the cases when velocities differ essentially from the ``spotlight velocity''.
It is also notable 
that such an effect can not be reached for the case where 
$ \alpha = 35^\circ $, 
$ \varepsilon = 4 $. 
For all other parameters indicated in Fig.~%
\ref{fig:5}, 
the ``spotlight velocity'' can be found and therefore the ``spotlight effect''can be realized.  

Note as well that approximate expressions~%
\eqref{eq:(53)} 
(for the case when 
$ \theta_t $ 
is not  small) and~%
\eqref{eq:(60)}, 
\eqref{eq:(62)} 
(for the case of ``Cherenkov spotlight'' when 
$ \theta_t = 0 $%
) give good coincidence with the results shown in Fig.~%
\ref{fig:5} 
(the discrepancy in the areas of the high field values does not exceed a few percent).

%%%%%%%%%%%%%%%%%%%%%%%%%%%%%%%%%%%%%%%%%%%%%%%%%%%%%%%%%%%%%%%%%
%%%%%%%%%%%%%%%%%%%%%%%%%%%%%%%%%%%%%%%%%%%%%%%%%%%%%%%%%%%%%%%%%
\section{\label{sec:concl} Conclusion }
%%%%%%%%%%%%%%%%%%%%%%%%%%%%%%%%%%%%%%%%%%%%%%%%%%%%%%%%%%%%%%%%%
%%%%%%%%%%%%%%%%%%%%%%%%%%%%%%%%%%%%%%%%%%%%%%%%%%%%%%%%%%%%%%%%%

We have studied the radiation generated by a charge moving in vacuum channel through the ``inverted'' dielectric cone assuming that the cone sizes are much larger compared to the wavelengths of interest. The wave field outside the target was calculated using the ``aperture method'' developed in our previous papers. 

It is worth noting that contrary to the problems considered earlier, here the wave which incidences directly on the aperture is not the main wave, while the wave once reflected from the lateral surface is much more important. 
We have obtained the analytical results  for CR outside the target (including the ray optics area and the most interesting Fraunhofer area) and analyzed significant physical effects. 

The most promising effect is the Cherenkov spotlight phenomenon which allows reaching essential enhancement of the CR intensity in the far-field region at certain selection of the problem parameters (the field in the main maximum can be increased approximately in $ \sqrt{ kb}$ times). It is important as well that for the ``inverted'' cone geometry, this effect can be realized for arbitrary charge velocity, including the case $ \beta \approx 1 $, by proper selection of the cone material and the apex angle.

%%%%%%%%%%%%%%%%%%%%%%%%%%%%%%%%%%%%%%%%%%%%%%%%%%%%%%%%%%%%%%%%%
%%%%%%%%%%%%%%%%%%%%%%%%%%%%%%%%%%%%%%%%%%%%%%%%%%%%%%%%%%%%%%%%%
\section{Acknowledgements}
%%%%%%%%%%%%%%%%%%%%%%%%%%%%%%%%%%%%%%%%%%%%%%%%%%%%%%%%%%%%%%%%%
%%%%%%%%%%%%%%%%%%%%%%%%%%%%%%%%%%%%%%%%%%%%%%%%%%%%%%%%%%%%%%%%%

This research was supported by the Russian Science Foundation, Grant No.~18-72-10137.

%\bibliography{SNG_Bibliography_Dec2019}
%

\end{document}